\documentclass[12pt]{article}

\usepackage{graphicx}

\newcommand{\be}{\begin{equation}}
\newcommand{\bea}{\begin{eqnarray}}
\newcommand{\ee}{\end{equation}}
\newcommand{\eea}{\end{eqnarray}}

\def\x{\mbox{{\bf x}}}
\def\g{\mbox{{\bf g}}}
\def\R{\mbox{{\bf R}}}
\def\Y{\mbox{$\bf{Y}$}}
\def\X{\mbox{$\bf{X}$}}
\def\y{\mbox{$\bf{y}$}}
\def\w{\mbox{$\bf{w}$}}
\def\p{\mbox{$\bf{p}$}}
\def\F{\mbox{$\bf{F}$}}

\begin{document}

\begin{center}
\bigskip
\bigskip
\bigskip
{\bf {\Large State and Parameter Estimation using Monte Carlo Evaluation of Path Integrals}}\\
\bigskip
\bigskip
\bigskip
\bigskip
\bigskip
J.C. Quinn\\
\bigskip
Department of Physics\\
\bigskip
and\\
\bigskip
Henry D. I Abarbanel\\

\bigskip

Department of Physics\\ and\\ Marine Physical Laboratory (Scripps Institution of Oceanography)\\ 
Center for Theoretical Biological Physics\\
\bigskip
\bigskip
\bigskip
\bigskip
University of California,
San Diego,\\ 9500 Gilman Drive, Mailcode 0402,\\ La Jolla, CA
92093-0402   USA\\ 
\bigskip
\bigskip
\today 
\end{center}

\clearpage 

\begin{abstract}
Transferring information from observations of a dynamical system to estimate the fixed parameters and unobserved states of a system model can be formulated as the evaluation of a discrete time path integral in model state space. The observations serve as a guiding potential working with the dynamical rules of the model to direct system orbits in state space. The path integral representation permits direct numerical evaluation of the conditional mean path through the state space as well as conditional moments about this mean. Using a Monte Carlo method for selecting paths through state space we show how these moments can be evaluated and demonstrate in an interesting model system the explicit influence of the role of transfer of information from the observations. We address the question of how many observations are required to estimate the unobserved state variables, and we examine the assumptions of Gaussianity of the underlying conditional probability.
\end{abstract}

\section{Introduction} 

Using observations to estimate unknown parameters and unobserved state variables in a model of a dynamical system when the measurements are noisy, when the model has errors, and when the initial state of the system is uncertain is a challenge across many areas of scientific inquiry. The goals in the estimation procedure might focus only on the knowledge of fixed parameters to characterize a model and allow, with a new initial state, the prediction of a response to new forcing or stimuli, or one may wish to know both the parameters and the full state of the model system at the end of data acquisition to permit prediction from that point forward in time. 

If the model system, and presumably the physical system it represents, expresses chaotic orbits, then repeated acquisition of new data to inform it of the correct location in its state space is likely required. Absent observations in occasional temporal windows, the positive Lyapunov exponents of the model system will cause errors in the state variables at the end of the assimilation window to grow exponentially rapidly, leading to the loss of all predictability of a specific orbit. Forecasts in geophysical settings often fall into this category.

We have not made a thorough survey of the arenas where state and parameter estimation problems are important, but we have examined specific papers estimating parameters and states in neurobiology~\cite{huys,huys2}, systems biology~\cite{rod}, atmospheric and oceanic sciences~\cite{annan,lorenc}, toxicology~\cite{lyons} biomedical 
engineering~\cite{horv}, cell biology~\cite{beard,panning}, chemical engineering~\cite{xiong}, coastal and estuarine modeling~\cite{yang}, wastewater treatment~\cite{mueller}, biochemistry~\cite{dochain}, 
and immunology~\cite{swam} as examples. Constructing observers in control theory~\cite{nij1} also deals with state estimation from observed data. 

This problem requires the evaluation of a discrete time path integral~\cite{abar09} through the state space of the model system where measurements act as a `potential' transferring information to the model and guiding it toward the measured state and away from phase space locations where the model chaotic behavior might carry it.

\section{Path Integral Formulation}

We start with observations of $L$ quantities $\y(n) = \{y_1(t_n), y_2(t_n), ..., y_L(t_n)\}$ made in an observation window at the discrete times $t_n = \{t_0, t_1, ..., t_m\}$. These are known functions $h_l(\w(t)); l = 1, 2, ..., L$ of the state $\w(t)$ of the observed system. We construct a model of this system in D-dimensional space $\x(t_n) = \x(n) = \{x_1(n), x_2(n), ..., x_L(n), x_{L+1}(n), ..., x_D(n) \}$ and associate the first $L$ components of $\x(t_n) = \{x_1(n), ..., x_L(n)\}$ with the measurements $y_l(n); l = 1, 2, ..., L$~\cite{abar09}. The other $D-L$ state variables are unobserved. The model has time-independent parameters $\p$, including those inherited from the measurement functions $h_l$. The model dynamics is constructed on the basis of physical or biophysical reasoning not discussed here, and it is Markov: the state $\x(n+1)$ is given in terms of the state $\x(n)$ and the $\p$ via the 
dynamical rule $g_a(\x(n),\x(n+1),\p) = 0;\;a=1,2,...,D;\; n = 0, 1, 2, ..., m-1$.

In the situation where measurements are noisy, the model has errors, and the initial state of the system is uncertain, we seek to estimate the conditional probability $P(\x(m)|\Y(m))$~\cite{jaz,gordon} that the model state at time $t_m$ is at $\x(m)$, given the observations 
$\Y(m) = \{\y(m),\y(m-1), ..., \y(0)\}$. Using Bayes' rule~\cite{bayes} and the Chapman-Kolmogorov equation, both identities, we have the exact recursion relation 
\bea
&&P(\x(m)|\Y(m)) = \exp[MI(\x(m),\y(m)|\Y(m-1))] \nonumber \\
&& \int d^D x(m-1) P(\x(m))|\x(m-1)) P(\x(m-1)|\Y(m-1)),
\eea
where the conditional mutual information between the model state $\x(m)$ and the observations $\y(m)$, given the previous observations $\Y(m-1)$, is~\cite{fano}
\be 
MI(\x(m),\y(m)|\Y(m-1)) = \log \biggl[\frac{P(\x(m),\y(m)|\Y(m-1))}{P(\x(m)|\Y(m-1))\,P(\y(m)|\Y(m-1))}\biggr],
\ee
and $P(\x(n+1))|\x(n))$ is the transition probability to go from $\x(n)$ at $t_n$ to $\x(n+1)$ at $t_{n+1}$. 

Iterating back from $t_m$ to the beginning of the observation window at $t_0$ gives us $P(\x(m)|\Y(m))$ in terms of a discrete time path integral along paths $\X = \{\x(m),\x(m-1),...,\x(0)\}$ through $m+1$ time points in D-dimensional space:
\be 
P(\x(m)|\Y(m)) = \int \prod_{n=0}^{m-1}  d^Dx(n) \exp[-A_0(\X,\Y(m))],
\ee
where the action $A_0(\X,\Y(m))$ is given as
\be 
A_0(\X,\Y(m)) = -\sum_{n=0}^m MI(\x(n),\y(n)|\Y(n-1)) - \sum_{n=0}^{m-1} \log P(\x(n+1)|\x(n)) - \log P(\x(0)).
\ee
The first term is the total conditional mutual information transferred from the observations to the model. The final term reflects uncertainty in the initial state at the beginning of the assimilation window.

The conditional expectation value of a function $\chi(\X)$ on the path $\X$ is given by 
\bea 
E[\chi(\X)|\Y(m)] =\, <\chi(\X)> &=& \frac{\int \prod_{n=0}^m d^Dx(n)  \chi(\X) \exp[-A_0(\X,\Y(m))]}{\int \prod_{n=0}^m d^Dx(n) \exp[-A_0(\X,\Y(m))]} \nonumber \\
&& \nonumber \\
&=& \frac{\int d\X \chi(\X) \exp[-A_0(\X,\Y(m))]}{\int d\X \exp[-A_0(\X,\Y(m))]} 
\label{depi}
\eea
The contributions of fluctuations along the path $\X$ are accounted for through the integral. We call this the dynamical estimation path integral (DEPI). There are no approximations made in the formulation of this discrete time path integral.

In our considerations we are interested in the first through fourth moments of the components of $\X$, namely, the $[x_a(n)]^p;\;a=1,2,...,D;\,n=0,1,...,m;\;p=1,2,3,4$. This allows us to estimate the conditional mean trajectory, the RMS variation about it, as well as the skewness and kurtosis of the conditional distribution along the paths. This permits us to address the familiar assumption about the Gaussian nature of the conditional distribution where the skewness
\be 
\frac{<(x_a(n) - <x_a(n)>)^3>}{<(x_a(n) - <x_a(n)>)^2>^{3/2}},
\ee
and kurtosis
\be 
\frac{<(x_a(n) - <x_a(n)>)^4>}{<(x_a(n) - <x_a(n)>)^2>^2} -3
\ee
both vanish.

In the limit where time is continuous, this problem has been formulated using approximations to the action 
$A_0(\X,\Y(m))$~\cite{hochberg, eyink05,restrepo, apte07}, while in discrete time with similar approximations it has been considered as a noise reduction method~\cite{bpo}. 
We inherit one result from the analysis of continuous time: the temporal discretization of the noisy model equations 
\be 
\frac{d\x(t)}{dt} = \F(\x(t),\p) + \eta(t)
\ee
should satisfy ($\x(n) = \x(t_n=t_0 + n\Delta t)$) for small $\Delta t$:
\be 
\frac{\x(n+1) -\x(n)}{\Delta t} = \frac{\F(\x(n+1),\p) + \F(\x(n),\p)}{2} + \eta(\frac{t_n+t_{n+1}}{2}),
\ee
where $\eta(t)$ is the noise term (or model error term), and $\Delta t$ is the time step. This gives us the 
physically~\cite{hochberg} correct form of the dynamics for small $\Delta t$:
\be 
g_a(\x(n),\x(n+1),\p) = \x(n+1) - \x(n) - \Delta t \frac{\F(\x(n+1),\p) + \F(\x(n),\p)}{2}
\ee 
to use in the Markov transition probability $P(\x(n+1)|\x(n))$. This result harks back at least to~\cite{pythian,jouvet}, and perhaps further.

In an earlier paper~\cite{abar09} we discussed approximations to DEPI through the introduction of an 
effective action~\cite{zinn}. Loop expansions and renormalization group approximations to the integrals Eq. (\ref{depi}) are standard in statistical physics. Here we wish to use many common assumptions about the structure of the action $A_0(\X,\Y(m))$ to see how the use of the path integral can reveal information about the orbits of a chaotic nonlinear system provided with occasional observations. The problem, fundamentally, whatever approximations one uses, is to evaluate a high dimensional integral of dimension $(m+1)D + K$ where we have $K$ fixed parameters to estimate as well as D state variables at m+1 time locations. We use Monte Carlo methods for this purpose.

The literature on Monte Carlo methods is vast and interesting~\cite{neal,hastings,rossky,gamerman,rubin,duane,mack,roberts}. We have explored many different formulations of that technique, but here we report on the use of the standard Metropolis-Hastings approach.  A series of paths $\X^{i} ,\, i=1,2,...,N$ are generated according to the distribution $\exp[-A_0(\X,\Y(m))]$, and are then used to approximate Eq.~(\ref{depi}) as 
\be
<\chi(\X)> \approx \frac{1}{N} \sum_{i=1}^{N} \chi(\X^i)
\ee

\section{Approximations to $A_0(\X,\Y(m))$}

To proceed we must make some approximations to the terms in $A_0(\X)$. (We no longer display the 
measurements $\Y(m)$.)  We usually select the familiar assumption that the measurement 
noise is Gaussian at each time $t_n$, though later we examine another noise distribution as well, and we assume that measurement errors are uncorrelated at different times. As noted 
in~\cite{hamill} there may be important physical settings where this assumption is incorrect. This situation would demand another approximation to the total conditional mutual information term in the action.

For independent, Gaussian observation errors at each observation time we write for the first term in $A_0(\X)$
\be 
-\sum_{n=0}^m MI(\x(n),\y(n)|\Y(n-1)) = \frac{1}{2}\,\sum_{n=0}^m \sum_{l,l'=1}^L (y_l(n) - x_l(n))(\R_m)_{ll'}(y_{l'}(n) - x_{l'}(n)),
\ee
and, if the errors of an observation are independent of errors in other observations, then the $L \times L$ matrix $\R_m$ is diagonal. We go even further for illustration purposes and consider the $L \times L$ matrix $\R_m$ to be a multiple of the identity; this means the errors in various components of the observation vector are all of the same magnitude. Our approximation to the total conditional mutual information term in $A_0(\X)$ takes the form
\be 
-\sum_{n=0}^m MI(\x(n),\y(n)|\Y(n-1)) = \frac{R_m}{2}\,\sum_{n=0}^m \sum_{l=1}^L (y_l(n) - x_l(n))^2,
\label{GaussianMI}
\ee
with $R_m$ a scalar.

In the case of no model error and perfect model resolution, one would write
\be 
P(\x(n+1)|\x(n)) = \delta^D(\g(\x(n),\x(n+1),\p)).
\ee
If we have resolution $\sigma_f$ in the model, this broadens the delta function to a smoother distribution, one representation of which is
\be 
\delta^D(\g(\x(n),\x(n+1),\p)) \to \sqrt{\frac{1}{(2\pi)^{D}\sigma_f^2}}\exp[-\frac{\g(\x(n),\x(n+1),\p)^2}{2\sigma_f^2}].
\ee
This reduces to the delta function as $\sigma_f \to 0$. This is certainly a minimalist representation of error in deterministic models, and it can hardly account properly for terms in the model being absent, for example, but it may give a sense of the effect of environmental noise as it impairs model output resolution. While there are more thoughtful discussions of model error~\cite{hansen,tremolet,judd08}, this simple assumption on model resolution will allow us to interpret the role of model error in assimilating information from data.

The final term in the action is related to the initial distribution of the state variable at the start of data assimilation $P(\y(0))$. 
We report here on calculations made assuming the initial conditions are distributed uniformly. This distribution has essentially no information about the orbits in state space, and one can, of course, do better with some knowledge of the state of the system at the beginning of data assimilation. In approaches such as the ensemble Kalman filter~\cite{evensen,kalnay1,kalnay2} this information is placed in a mean or background state estimate at $t_0$ and a covariance about this mean. 
Our experience showed that the initial distribution mattered very little as the dynamics takes the orbit from each initial condition and moves it onto or near the dynamical attractor because of the dissipation and the resulting strange attractor in the dynamics $g_a(\x(n+1),\x(n),\p) = 0$ as suggested by the arguments of~\cite{hansen}.

Our action then has two overall constants $R_m$ and $R_f = \frac{1}{\sigma_f^2}$ as well as any parameters that enter the dynamical model $\g(\x(n),\x(n+1),\p) = 0$. By considering the parameters as state `variables' satisfying $\p(n+1) = \p(n)$, we incorporate them into any estimation protocol we utilize.

\section{Evaluation of the Path Integral for the Lorenz96 Model}
We have examined the DEPI method using the model of Lorenz~\cite{lorenz96,emm} with $D = 20$ degrees of freedom for a range of $R_m$ and $R_f$ values. The dynamical equations for this model are
\be
\frac{dw_a(t)}{dt} = w_{a-1}(t)(w_{a+1}(t) - w_{a-2}(t)) - w_a(t) + f ;\; a=1,2,...,D
\label{lor96eqn}
\ee
with $w_{-1}(t) = w_{D-1}(t), w_0(t) = w_{D}(t)$, and $w_{D+1}(t) = w_1(t),$ with $D$ = 20, some choice of $w_a(0)$ and with the forcing parameter selected to be $f = 8.17$ creating chaotic orbits $\w(t)$. We generated a solution to the dynamical equations and added Gaussian white noise to the $w_a(t_n)$ to act as our measurements: $\y(n) = \w(n) + \mbox{noise}$. 

This is a twin experiment in which the model equations are (\ref{lor96eqn}) for variables $x_a(t)$. We evaluated the first through fourth moments of the paths $\X$ so we could estimate the conditional mean $<x_a(n)>;\;a=1,2,...D; n=0,1,2,...m$ as well as the RMS variation about this mean and the skewness and kurtosis associated with the  distribution of $\x(n)$ at the temporal locations along the path. The latter tells us how the conditional probability differs from a Gaussian within and without any assimilation windows.

The time step in solving these equations was $\Delta t = 0.05$~\cite{lorenz96,emm} corresponding to 6 hours in physical time. We selected paths from a space of dimension $(m+1)D + K$ with $m = 80$ time steps, $D = 20$ state variables and $K = 1$ parameters.  We started from an arbitrary initial path and ignored the first $3 \times 10^5$ paths generated, and then used the next $1.2 \times 10^6$ paths for the evaluation of the moments of $\X$.  The computation time per path generated will depend on the number of terms in the action that need to be re-computed when a single component of the path is updated.  For the Lorenz96 model the number of terms that need to be computed to update the entire path is $(8+K)(m+1)D$.  

\subsection{How many Observations are Required?}\label{A0shape}
It is necessary to decide how many observations are required to allow the estimation of the {\tt unobserved} state variables.

\begin{figure}[ht]
\hspace{-0.5 in}
\includegraphics[width= 3.7in,angle=0]{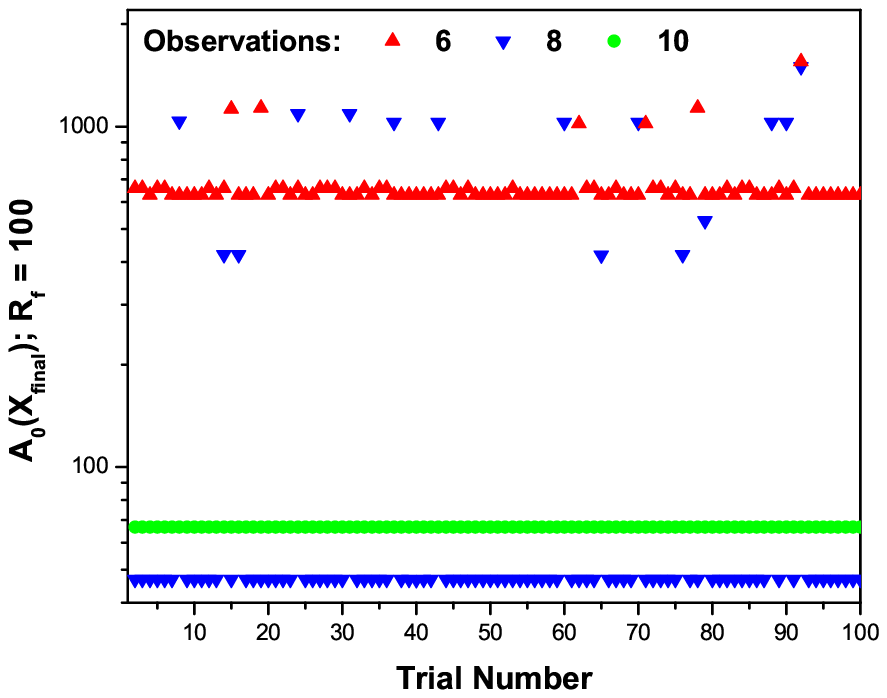}
\hspace{-0.2in}
\includegraphics[width= 3.7in,angle=0]{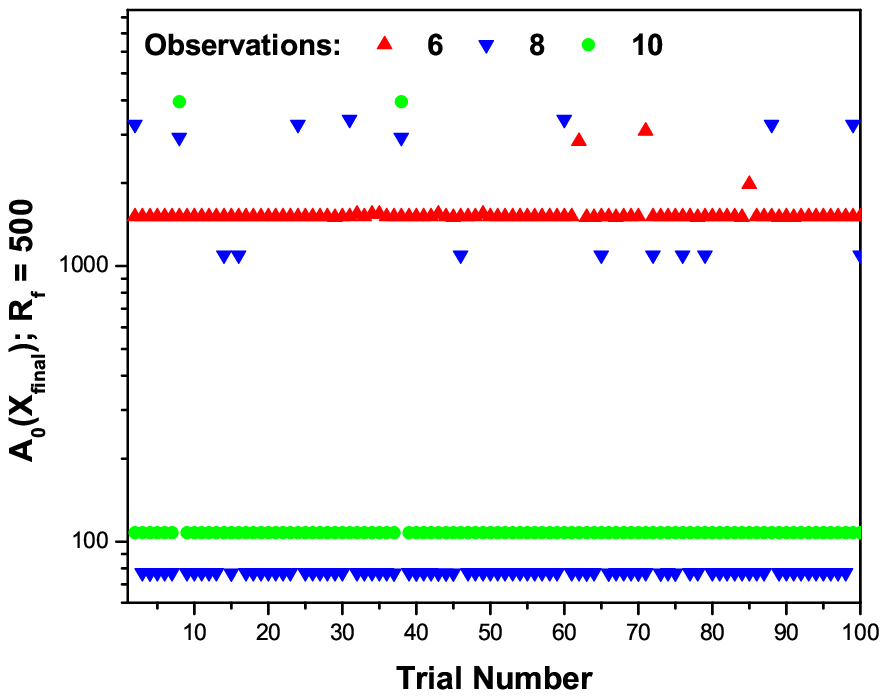}
\caption{Lorenz96, D = 20. $A_0(\X_{final})$ for 100 initial choices of $\X(s=0)$ allowing the paths to evolve through Equation (\ref{langevin}), with the number of observed variables $L = 6, 8,$ or 10. {\bf Left}: $R_f = 100$. {\bf Right}: $R_f = 500$.}
\label{lor9620a0}
\end{figure}

We know that as $\sigma_f \to 0$ or $R_f$ becomes very large, we reach the deterministic, no model-error, setting. In this limit the search over state and parameter values becomes a search in the space of initial conditions and parameters with dimension $D + K$, and is impeded by complex surfaces with many local minima in $A_0(\X)$ associated with instabilities on the synchronization manifold~\cite{acg,abarsiads} $x_l(n) \approx y_l(n);\;l=1,2,...,L$. In our case where $R_f$ is finite, we still see the remnants of this instability in the higher dimensional space of $(m+1)D + K$ dimensions, but it is much less of an issue. In 
Figure (\ref{lor9620a0}) we display the action $A_0(\X_{final})$ evaluated by solving the differential equation
\be 
\frac{d\X(s)}{ds} = -\frac{\partial A_0(\X(s))}{\partial \X(s)},
\label{langevin}
\ee
for 100 random selections of $\X(s=0)$ using the $A_0(\X)$ associated with the discrete time formulation of the Lorenz96, D = 20 model. This `Lang\'evin' equation (without noise here) has $A_0(\X)$ as a Lyapunov function, since
\be 
\frac{dA_0(\X(s))}{ds} = - \biggl (\frac{d \X(s)}{ds}\biggr )^2,
\ee
and converges to paths $\X_{final}$ at locations in $\X$ space where $\frac{\partial A_0(\X)}{\partial \X} = 0$. Equation (\ref{langevin}) is of interest because, if one adds Gaussian noise of amplitude $\sqrt{2}$ to the right hand side of this equation, the paths associated with that Lang\'evin equation are distributed in $\X$ space as 
$\exp[-A_0(\X)].$  

The distribution of values of the action at the locations $\X_{final}$ are shown for $L = 6, 8,$ and $10$ observations, $R_m = 50$, with $R_f = 100$ and with $R_f = 500$ in Figure (\ref{lor9620a0}). 
For the Monte Carlo approximation of the path integrals of Eq.~(\ref{depi}) to be effective, the most probable paths should all be clustered around a well defined global minimum of $A_0(\X)$.  Figure~(\ref{lor9620a0}) suggests that there are two ways within the DEPI approach to make the surface of $A_0(\X)$ smoother.  One way is to increase the number of observed variables.  This causes the number of local minima of $A_0(\X)$ to be reduced,  because information from measurements reduces the number of likely paths.  The other way is to increase the uncertainty of the model, by decreasing $R_f$.  

In Figure (\ref{lor9620a0}) we investigate increasing the number of observed variables. The figure shows that $L = 6$ observations are not enough, but $L = 8$ are enough to smooth the surface.  Even though there still some local minima for $L \ge 8$, the action around these minima is much larger than at the global minimum, so those paths contribute a negligible amount to the path integrals.

In the deterministic case, we know that the minimum number of observations required to remove the instabilities is equal to the number of positive conditional Lyapunov exponents (CLEs) associated with the synchronization manifold~\cite{acg,abarsiads} for the model equations when they receive the obervations $\y_l(t_n)$. We have examined this for the Lorenz96 model and found that about $0.4D$ observations are needed to make all 
CLEs negative~\cite{mark}. For this reason we selected $L = 8$ for our analysis of the $D = 20$ dimensional Lorenz96 system. In our calculations we chose to `observe' $y_a(n)$ for a = 0, 3, 5, 8, 10, 13, 15 and 18.  It is not necessary to have observations at every time step, and in fact here we only provide observations for even $n$.  The missing observation terms are excluded from the conditional mutual information contribution to the action.

We emphasize that this criterion for selecting the required number of observations depends on the model, which is very useful. It also depends on the accuracy of the model, here represented by $R_f$. The more accurate the model, namely the larger $R_f$, the more local minima there will be in $A_0(\X)$, because of instabilities on the synchronization manifold of the deterministic problem~\cite{acg,abarsiads}.

\subsection{Results of Monte Carlo Estimation of the Path Integral for Moments of $\X$}

We report here on example calculations using the Lorenz96 model with $D=20$ and 8 observed variables. We used a data assimilation window of 80 steps ($0 \le t \le 4$) with observations every other step, followed by a prediction window ($4 < t \le 6$) with no measurements.  ($\Delta t = 0.05$ as above.) In most cases we considered, the moment calculations of $\x(n)$ in the prediction window were computed in the deterministic limit ($R_f \rightarrow \infty$) by integrating the model equation forward in time using a 4th order Runge-Kutta procedure.  The integration was done by taking initial conditions $\x(t_m)$ and parameter values from each sampled path at the end of the assimilation window. This effectively evolves the whole conditional distribution at the end of the assimilation window forward in time into the prediction window.  We also did a simpler type of prediction for comparison, where the model equation is integrated forward using only the conditional mean values $<\x(t_m)>$ and $<\p>$ at the end of the data assimilation window as initial conditions and parameters respectively.  These two methods give quite comparable results, but the first provides more information about the distribution of states.

\subsubsection{Prediction by Model Equations for $t > t_m$}

\begin{figure}[htb]
\hspace{-0.5in}
\includegraphics[width= 3.5in]{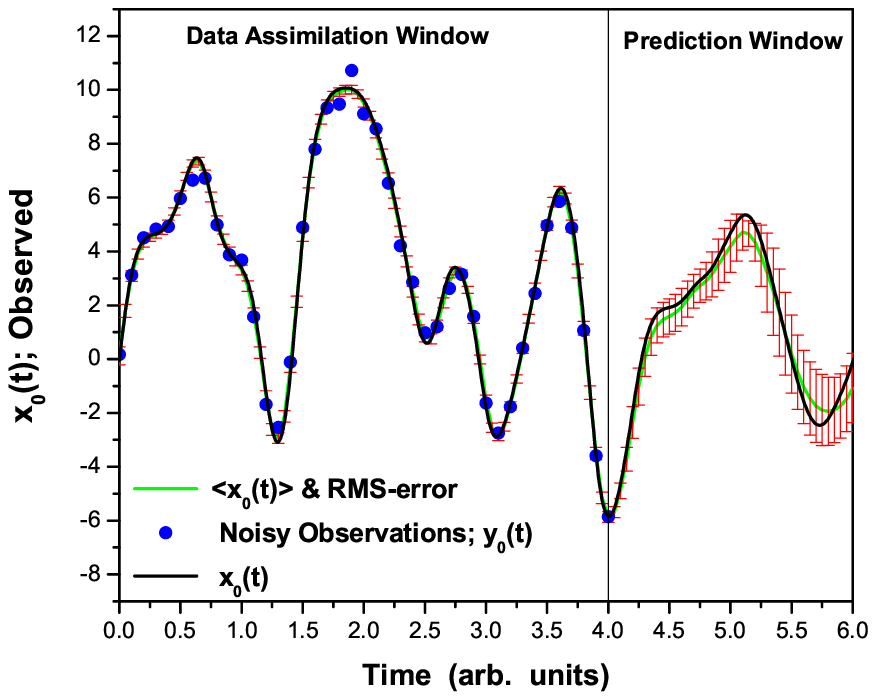}
\includegraphics[width= 3.5in]{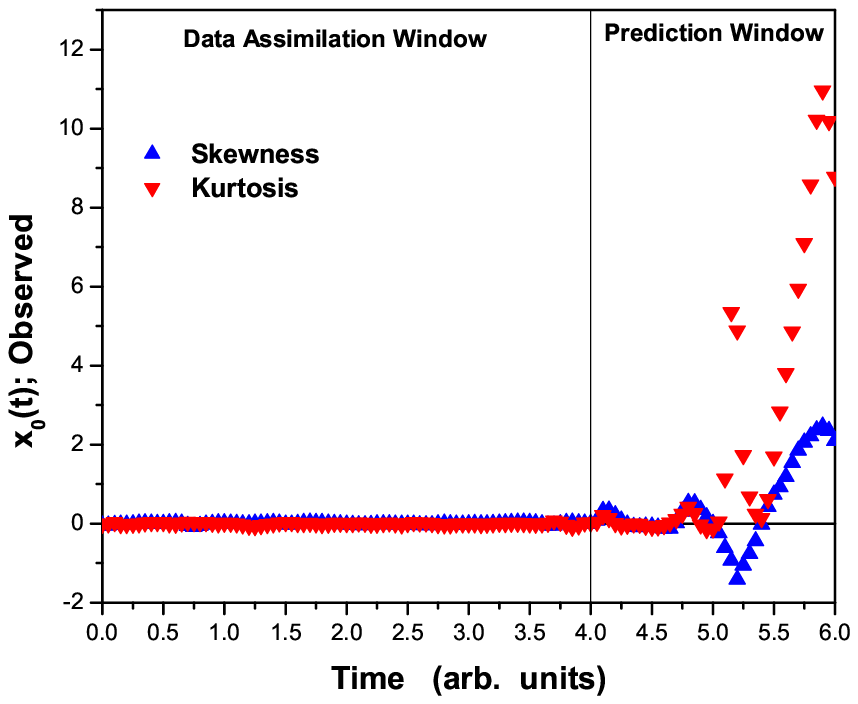}
\caption{Observed variable $x_0(t)$. {\bf Left}: Conditional Mean (Green) and RMS-error (Red) along with Gaussian distributed noisy observations (Blue) and known $x_0(t)$ (Black). {\bf Right}: Skewness (Blue) and Kurtosis (Red) of $x_0(t)$. These are Monte Carlo estimates from the path integral for the  Lorenz96 model with $D = 20$; $R_m = 8, R_f = 100$. The assimilation window is $0 \le t \le 4$, and the prediction window is $4 < t \le 6$.  The parameter estimate is $f = 8.25 \pm 0.09$. Predictions for $t > 4$ are made with a fourth order Runge-Kutta procedure using information on the parameter and state variables at $t = 4$.}
\label{x0rf100}
\end{figure}

\begin{figure}[htb]
\hspace{-0.5in}
\includegraphics[width= 3.5in]{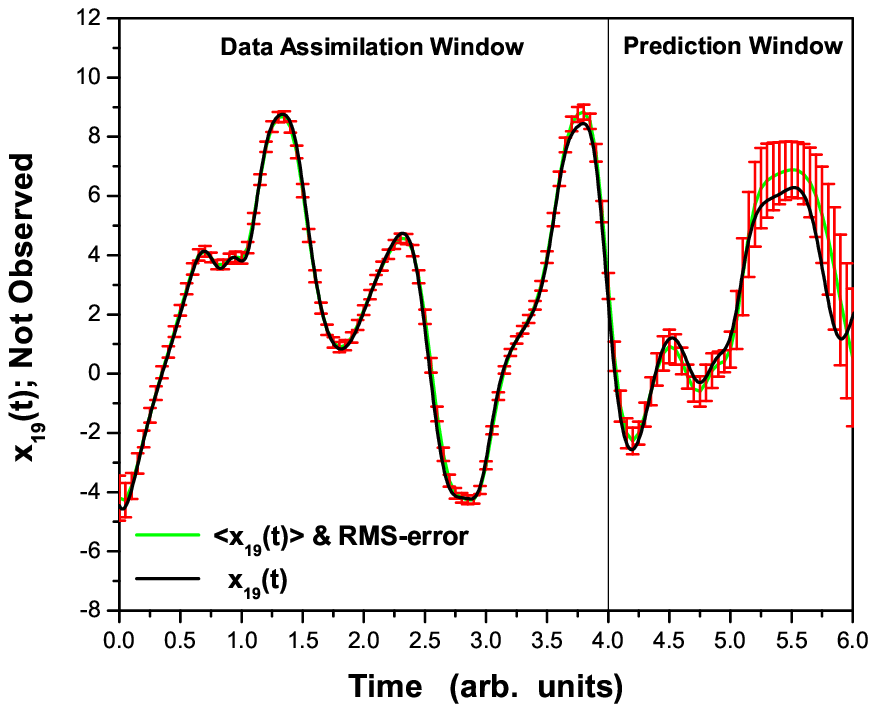}
\includegraphics[width= 3.5in]{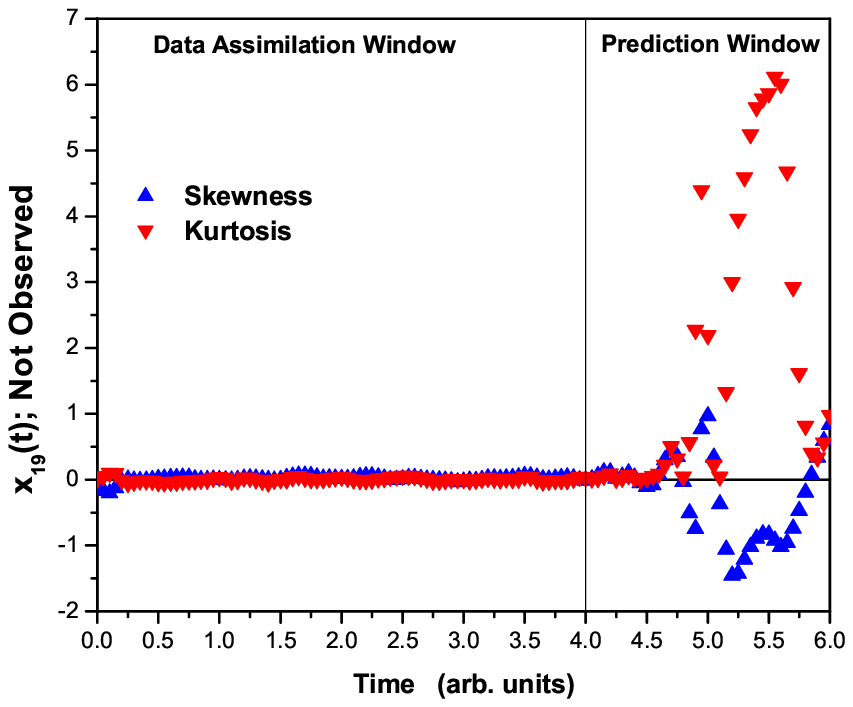}
\caption{Unobserved variable $x_{19}(t)$. {\bf Left}: Conditional Mean (Green) and RMS-error (Red) along with Gaussian distributed noisy observations (Blue) and known $x_{19}(t)$ (Black). {\bf Right}: Skewness (Blue) and Kurtosis (Red) of $x_{19}(t)$. These are Monte Carlo estimates from the path integral for the  Lorenz96 model with $D = 20$; $R_m = 8, R_f = 100$. The assimilation window is $0 \le t \le 4$, and the prediction window is $4 < t \le 6$.  The parameter estimate is $f = 8.25 \pm 0.09$. Predictions for $t > 4$ are made with a fourth order Runge-Kutta procedure using information on the parameter and state variables at $t = 4$.}
\label{x19rf100}
\end{figure}

The results of two of these calculations are shown in Figures~(\ref{x0rf100},\ref{x19rf100}) for two representative state variables, $x_0(t)$ (observed) and $x_{19}(t)$ (unobserved).  These behave in the same manner as the other eighteen variables, whether observed or not. We first created the `true path' (solid black lines) by integrating the model equations with $f = 8.17$ and some choice of initial conditions on the attractor.  This gives us all the state variables of the observed system. We then generated `observations' (blue dots) from eight of the state variables by adding Gaussian noise with standard deviation $\sigma_m = 0.353$ to the true path. There is no correlation in the noise at different time steps or among different variables.  

We selected $R_m = 8 \approx 1/\sigma_m^2 $ and $R_f =100$ for these calculations.  The estimated states (for $0\le t \le 4$) and the predicted states (for $4 < t \le 6$) are shown as green lines with red error bars representing the conditional mean plus or minus one standard deviation.  Also shown in the Figures are the skewness and kurtosis of the state variables at each time step.  

The state estimates track the true path quite well in the assimilation and prediction windows for the observed as well as the unobserved state variables.  The uncertainties of the predicted states grow in time, because the largest Lyapunov exponent of the model is positive, about 0.9 in units of inverse 
time~\cite{mark}. This means that at the end of the prediction period, $t=6$, the uncertainties should be about six times as large as the uncertainties at the end of the assimilation window, $t=4$.    

In the example of Figures~(\ref{x0rf100}) skewness and kurtosis are both close to zero in the assimilation window.  They are slightly smaller in magnitude for $R_f = 100$ than for $R_f= 500$ (not shown).  This suggests that the conditional distributions are nearly Gaussian during the assimilation window, probably because of the influence of the measurements.  The ratio 
$\frac{R_m}{R_f}$ is the determining factor. When this is sizeable, the Gaussian errors in the observations are important. When this ratio goes to zero, in the deterministic or zero model error limit, the non-Gaussian part of the action is dominant. When the assimilation window ends, the distribution is evolved according to the nonlinear dynamics of the model, and so it becomes much more non-Gaussian and less localized because of chaos.  The distribution can become quite complicated possibly with the regions containing the most probable paths no longer contiguous in path space, and so extending the Monte Carlo evaluation into the prediction window is likely to be difficult. Nonetheless we show below an example that attempts this.

This suggests that as the models of the observed process become better and better, namely, model error is reduced, the role of the nonlinear, non-Gaussian elements of the action $A_0(\X,\Y(m))$ will be more and more important. Approximations to the assimilation of information from measurements based on Gaussian assumptions may become less valuable in this circumstance.

\subsubsection{Prediction within the Path Integral}

\begin{figure}[htb]
\hspace{-0.5in}
\includegraphics[width= 3.5in]{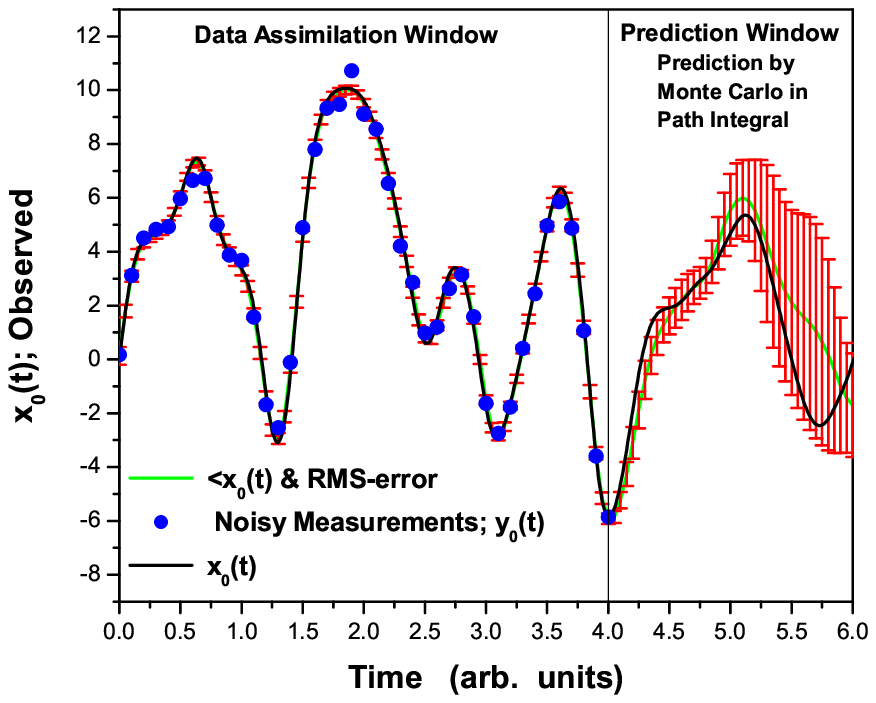}
\includegraphics[width= 3.5in]{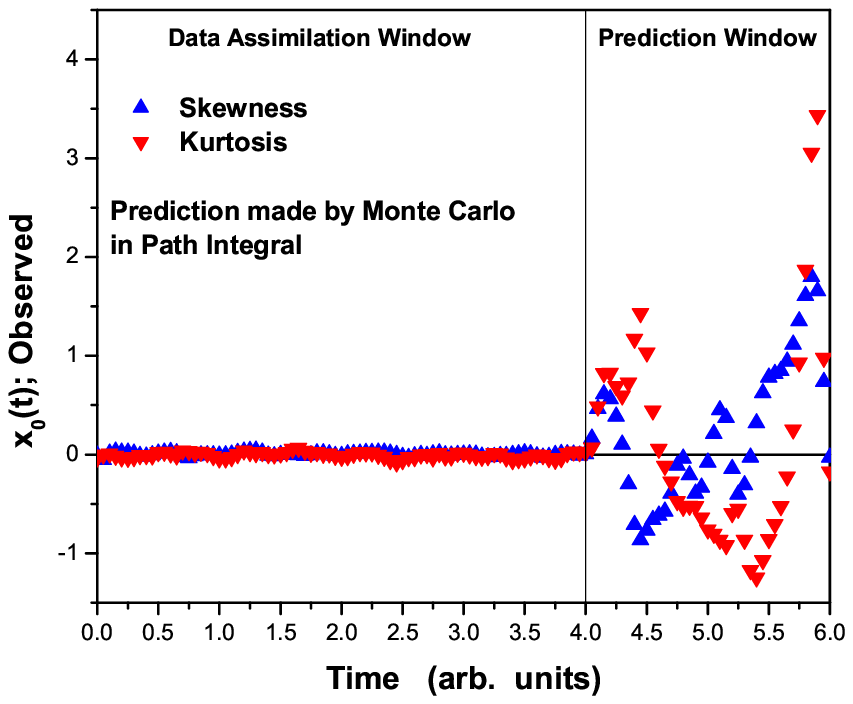}
\caption{Observed variable $x_{0}(t)$. {\bf Left}: Conditional Mean (Green) and RMS-error (Red) along with Gaussian distributed noisy observations (Blue) and known $x_{0}(t)$ (Black). {\bf Right}: Skewness (Blue) and Kurtosis (Red) of $x_{0}(t)$. These are Monte Carlo estimates from the path integral for the  Lorenz96 model with $D = 20$; $R_m = 8, R_f = 100$. The assimilation window is $0 \le t \le 4$, and the prediction window is $4 < t \le 6$.  The parameter estimate is $f = 8.24 \pm 0.09$. Predictions are made within the Monte Carlo evaluation of the path integral.}
\label{x0rf100mcprediction}
\end{figure}

We also performed a Monte Carlo path integral calculation to estimate the moments both within the data assimilation window and within the prediction window, and show the results in Figure (\ref{x0rf100mcprediction}). In this we select paths to evaluate the path integral over the whole time interval $0 \le t \le 6$ including both a data assimilation and a window with no observations. The decreased resolution of the model dynamics represented in the path integral by finite $R_f$ now plays a role in the quality of the predictions for $t > 4$. We see that the prediction is slightly worse than when we used the model equations for $t > 4$, while the deviation from Gaussianity in the data assimilation window is increased.

\subsubsection{Non-Gaussian Measurement Error}

In the use of the path integral method, or any other data assimilation approach actually, we do not know the statistics of the error in the measurements, and while the assumption that they are Gaussian is common, it is by no means necessary. To examine the implication of selecting another distribution of the errors, we represented the measurement errors by a Lorentzian distribution (also known as a Cauchy distribution)
\be 
P(z) \propto \frac{1}{(1+z^2)^4}.  
\ee

This replaces the conditional mutual information term in the action by 
\be 
-\sum_{n=0}^m MI(\x(n),\x(n)|\Y(n-1)) = 4\sum_{n=0}^m \sum_{l=1}^L \log \left(1 + \frac{R_m}{2} (y_l(n) - x_l(n))^2 \right),
\label{LorentzianMI}
\ee

\begin{figure}[htb]
\hspace{-0.5in}
\includegraphics[width= 3.5in]{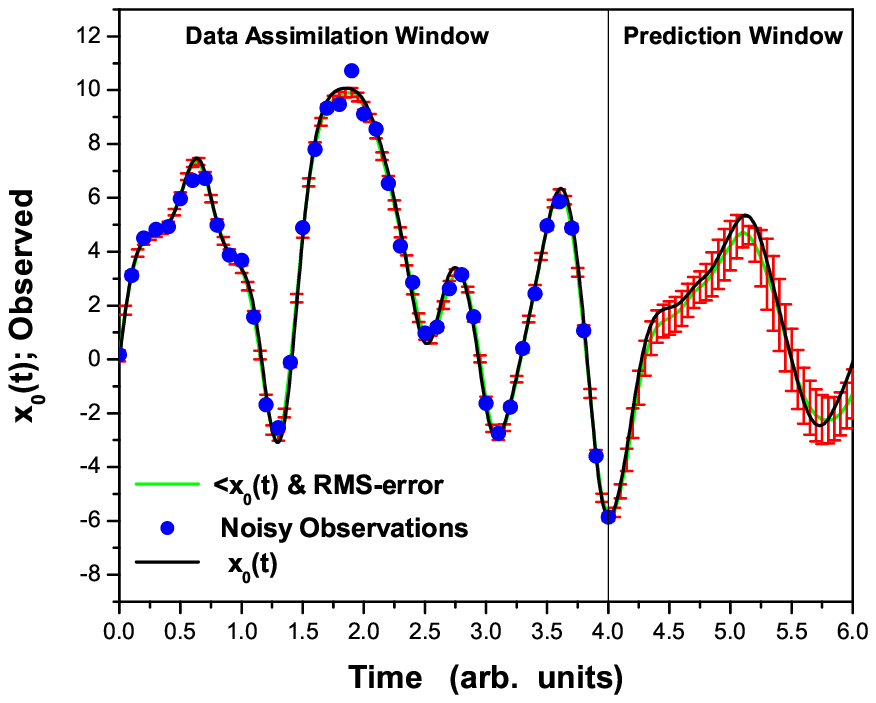}
\includegraphics[width= 3.5in]{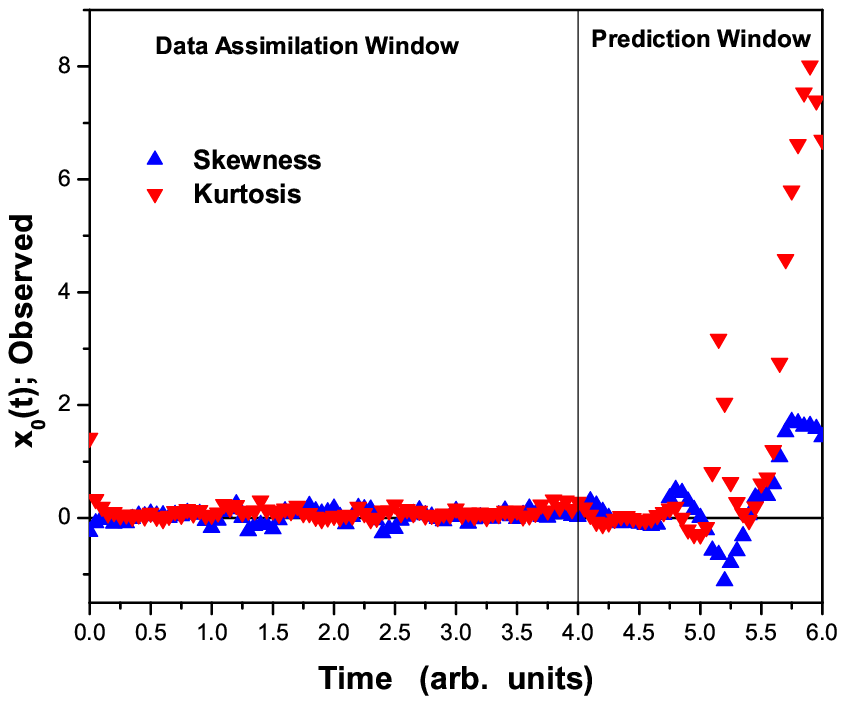}
\caption{Observed variable $x_{0}(t)$. {\bf Left}: Conditional Mean (Green) and RMS-error (Red) along with Gaussian distributed noisy observations (Blue) and known $x_{0}(t)$ (Black). {\bf Right}: Skewness (Blue) and Kurtosis (Red) of $x_{0}(t)$. These are Monte Carlo estimates from the path integral for the  Lorenz96 model with $D = 20$; $R_m = 8, R_f = 100$. The assimilation window is $0 \le t \le 4$, and the prediction window is $4 < t \le 6$.  In these calculations the conditional mutual information term in the action represents the measurement errors as Lorentzian, Equation (\ref{LorentzianMI}). The parameter estimate is $f = 8.24 \pm 0.08$. Predictions for $t > 4$ are made with a fourth order Runge-Kutta procedure using information on the parameter and state variables at $t = 4$.}
\label{x0rf100lorentzian}
\end{figure}

Figure~(\ref{x0rf100lorentzian}) shows the results of this calculation, using the same measurement data as the other two examples, and with $R_f = 100, R_m = 8$.  This change does not make a substantial difference for the values of $R_m$ and $R_f$ we used, except that the conditional distributions of the observed variables in the assimilation window are slightly less consistent with a Gaussian than before. The exploration of the effect of non-Gaussian measurement noise distributions can be accomplished in a straightforward manner through the use of DEPI.   

\subsubsection{Annealing as a Monte Carlo Tool}

Figure~(\ref{lor9620a0}) suggests that it might be useful to gradually decrease the model resolution during the initialization phase to avoid getting trapped in local minima and completely missing the global minimum.  We do this in all the examples by replacing $R_f$ by $\beta R_f$ in the action.  We start the simulated annealing process with $\beta = 0.01$ on the first iteration and end with $\beta = 1.00$ on the 200,000$^{th}$ iteration, by multiplying $\beta$ by a constant factor on every iteration, and then do another 100,000 initialization iterations before recording any path statistics.  We found that typically the paths near a local minima of $A_0(\X)$ differ from the true path at early times, but coincide at later times, because the dynamics is dissipative.  All paths contribute to the path integral, however these paths have significantly larger action than paths around the global minimum, and so have a negligible contribution because of the exponential weighting factor.
 
\section{Discussion}

We have presented the problem of incorporating information from noisy observations into model dynamics of the observed system as an exact discrete time path integral along orbits of the state variables of the model. The density of paths in the state space of the orbits is given by $\exp[-A_0(\X)]$ where $A_0(\X)$ is an action composed of terms conveying information from the measurements to the model, terms propagating the model between observations, and the distribution of states at the beginning of the temporal observation window. As expectation values of functions in the state space $\X$, $\chi(\X)$, are integrals over paths with this density, one is presented with high dimensional integrals to perform, and we have investigated the use of Monte Carlo methods for this purpose, after making simplifying assumptions about the elements of the action.

By selecting the function $\chi(\X)$ as the first through fourth powers of the state variables, we have evaluated the conditional mean state along the path as well as the RMS variation about the mean, and the skewness and kurtosis associated with fluctuations about that conditional mean. All moments are conditioned on the observations. As an example, much studied in the geophysical literature, we selected the model of Lorenz~\cite{lorenz96} with $D = 20$ state variables and one fixed, forcing parameter, working in a region of forcing where orbits of the model are chaotic. We performed a twin experiment wherein the data is generated by the model, noise is added to those orbits, and the model is used as a nonlinear filter through the path integral to estimate the value of the unobserved state variables as well as the forcing parameter. 

The number of observed states required to allow the estimation of the remaining states of the model system and any fixed parameters of the model is an important issue to be addressed, for with $L$ observations and a $D >>L$ dimensional model as is typical, one needs some sense of $L$ in order to proceed. In the case where there is no model error and the dynamics has perfect resolution, we know the number of pieces of information required from the observations should be, at minimum, the number of conditional Lyapunov exponents of the nonlinear dynamics of the model. When one has model errors and reduced resolution in the model state space, we showed that one may use properties of the action to estimate how many observations are required.

The tool we proposed is to look for minima in the action by solving the `Lang\'evin' equation for the paths as they evolve in ``time'' called $s$:
\be 
\frac{d\X(s)}{ds} = -\frac{\partial A_0(\X(s))}{\partial \X(s)},
\ee
starting at a selection of initial values $\X(s=0)$ and locating the minima associated with the final element to which this moves in $s$. For too few observations, there are many local minima, reflecting the complex structure of $A_0(\X)$ in state and parameter space associated with the instability of the manifold where the model output synchronizes with the observations~\cite{acg,abarsiads}. Adding measurements smoothes out the surfaces explored by this `time' evolution when model errors or diminished model resolution is present.

The use of this tool is suggested by the fact that adding Gaussian noise to the Lang\'evin equation leads to the density of paths $\exp[-A_0(\X)]$ required for the path integral. The examination of the minima of $A_0(\X)$ is essentially the four dimensional variational principle 4DVAR~\cite{lorenc} in the context of the path integral~\cite{abar09}. The path integral allows the evaluation of corrections due to fluctuations about this `optimal' path.

We then used a quite standard Monte Carlo Metropolis-Hastings~\cite{neal,hastings} method to select paths for evaluation of the moments desired. This is surely not the most computationally efficient Monte-Carlo approach, and, indeed, using properties of the Lang\'evin equation to select paths may be much more efficient~\cite{rossky,roberts}. 

We showed that in the example of the $D = 20$ dimensional Lorenz96 model, we were able to accurately estimate both the unobserved model state variables and the fixed forcing parameter with Monte Carlo methods applied to the path integral. We predicted the development of the model states after the data assimilation window closed by using the states and parameter at the end of this window as initial conditions in the deterministic model equations and as part of the continued use of the path integral itself. The former method gave quite accurate forecasts limited by the natural chaotic behavior of the model system which enhances any error in the estimates at the end of the assimilation window. The growth of this error in the deterministic model forecast is consistent with the known values of the largest Lyapunov exponent of the Lorenz96 model.

When we used the path integral to make forecasts after the data assimilation window closed, we were successful again in those predictions, though less so than when using the deterministic model equations. As we introduced explicit loss of model resolution into the path integral formulation, this should not be a surprise. This use of the path integral for both assimilation of observed information and forecasting may prove valuable.

As a final calculation we investigated the assumption of Gaussian measurement errors in the formulation of the path integral. We used our `data' generated by adding Gaussian noise to the model output, and assumed that in the formulation of the path integral the conditional mutual information term was represented by a Lorentzian distribution of measurement errors. This had very little effect on the conditional means of the state variables and forcing parameter. The calculated deviations from Gaussian distributions of the estimated state variables, represented by nonzero skewness and kurtosis, were larger in this setting than when Gaussian measurement errors were assumed, but not significantly so.

By examining the skewness and kurtosis of the estimates of the state variables we conclude that even though the model dynamics is both non-Gaussian and chaotic, during the assimilation period the deviations from Gaussianity may remain small when the relative weight of the conditional mutual information, represented by our parameter $R_m$ remains large enough relative to the representative of model error, called $R_f$ here. When $R_f$ is increased relative to $R_m$, the non-Gaussian contributions of model error to the action increase, and larger contributions to the estimated skewness and kurtosis result.
This is especially clear from our results in the prediction windows. When we use the deterministic model equations to predict, we effectively set $R_f \to \infty$, and it is clear that the skewness and kurtosis grow rapidly as the orbits move to the strange attractor of the system. This is also true when we use the path integral, as in Figure (4), to provide the moments in the prediction window.

We conclude from these investigations that one can determine when assumptions about the Gaussianity of the conditional distribution of state variables might be a good approximation, and in those situations the use of Kalman like filtering approaches can be productive~\cite{evensen,kalnay1,kalnay2}. Before assuming statistics one might do well to use our results to examine the situation for a given model, set of parameters, and collection of observations. As one improves the model resolution and the representation of physical processes, one can expect such linear approaches to become inadequate.

\section*{Acknowledgements}
We would like to acknowledge useful discussions with Chris Schroeder, Julius Kuti about Monte Carlo methods, and an extensive interaction with Eugenia Kalnay and Brian Hunt about Kalman filtering methods for these problems. Mark Kostuk was kind enough to provide us with the Lyapunov exponents of the Lorenz96 model. We also appreciate the hospitality of Dan Margoliash at the University of Chicago where one of us (HDIA) spent a sabbatical quarter while this work was performed. We have been supported during this work by National Science Foundation Grant \#IOS-0905076.

\end{document}